\documentclass[twoside]{article}
\usepackage{Proc_NTSE_18}

\newcommand{\beq}{\begin{equation}}
\newcommand{\eeq}{\end{equation}}
\newcommand{\beqa}{\begin{eqnarray}}
\newcommand{\eeqa}{\end{eqnarray}}

\begin{document}
\thispagestyle{plain}
\publref{myfilename}

\begin{center}
{\Large \bf \strut
  Towards high-precision nuclear forces from chiral effective field theory
\strut}\\
\vspace{10mm}
{\large \bf 
Evgeny Epelbaum$^{a}$}
\end{center}

\noindent{
\small $^a$\it Ruhr-Universit\"at Bochum, Fakult\"at f\"ur Physik und
        Astronomie, Institut f\"ur Theoretische Physik II,  D-44780 Bochum, Germany} 

\markboth{
E.~Epelbaum}
{
Towards high-precision nuclear forces} 

\begin{abstract}
  Chiral effective field theory is being developed into a precision
  tool for low-energy nuclear physics. I review the state of the
  art in the two-nucleon sector,  discuss applications to few-nucleon systems and address challenges that will have to be faced over the
  coming years. 
\\[\baselineskip] 
\end{abstract}

\section{Introduction}
\label{sec1}

Chiral effective field theory (EFT) is widely applied to
study low-energy structure and dynamics of nuclear systems. The method
relies on the spontaneously broken approximate SU(2)$\times$SU(2)
chiral symmetry of QCD and allows one to compute the scattering amplitude of pions,
the Nambu-Goldstone bosons of the spontaneously broken axial
generators, with themselves and with matter fields via a
perturbative expansion in momenta and quark masses, commonly 
referred to as the chiral expansion \cite{Bernard:1995dp}. The appealing features of this
method lie in its systematic and universal nature,
which allows one to establish model-independent perturbative
relations between low-energy observables in the Goldstone-boson and
single-baryon sectors and low-energy constants (LECs) of the effective
chiral Lagrangian.

When applied to self-bound systems such as atomic
nuclei, the method outlined above has to be modified appropriately to
account for the non-perturbative nature of the problem at hand by
resumming certain parts of the scattering amplitude.
According to Weinberg \cite{Weinberg:1990rz}, the breakdown
of perturbation theory is attributed to the appearance of
time-ordered diagrams, which would be infrared divergent in the limit
of infinitely heavy nucleons and are enhanced relative to their
expected chiral order. The quantum mechanical
Schr\"odinger equation provides a simple and natural framework
to resum such enhanced contributions to the $A \geq 2$-nucleon scattering amplitude
as it can be efficiently dealt with using a variety of available {\it ab initio}
continuum methods
\cite{Gloeckle:1995jg,Barrett:2013nh,Hagen:2012fb,Hergert:2012nb,Soma:2012zd,Lovato:2013cua},
see the contributions of Petr Navratil \cite{Petr} and James Vary \cite{James},
or lattice simulations \cite{Lee:2008fa,Lee:2016fhn,UGMLattice}, see the
contributions by 
Ulf-G.~Mei{\ss}ner \cite{Ulf} and Dean
Lee \cite{Dean} for selected highlights and exciting new developments
along this line. The
problem thus essentially reduces to the derivation of the 
nuclear forces and currents, defined in terms of the corresponding
irreducible (i.e.~non-iterative) parts of the amplitude, that are
not affected by the above-mentioned enhancement and can be worked out 
order by order in the chiral expansion. The resulting framework,
firmly rooted in the symmetries of QCD, allows one to derive
consistent nuclear forces and currents and
offers a systematically
improvable theoretical approach to few- and many-nucleon systems
\cite{Epelbaum:2008ga,Epelbaum:2012vx,Machleidt:2011zz}. 

Contrary to on-shell scattering amplitudes, nuclear forces and
currents are scheme-dependent quantities, which are affected by
unitary transformations or, equivalently, by nonlinear redefinitions
of the nucleon field operators. Care is therefore required to maintain
consistency between two- and many-nucleon forces and the current
operators, see section \ref{sec5} for a discussion. The unitary
ambiguity in the resulting nuclear potentials is significantly reduced (but not
completely eliminated) by the requirement of their
renormalizability.\footnote{Contrary to the on-shell S-matrix, 
  loop contributions to the nuclear forces and currents may 
  contain ultraviolet divergences which cannot be absorbed into the counterterms
of the effective Lagrangian, see section \ref{sec5}. When calculating the scattering
amplitude, such divergences cancel against the ones 
generated by ladder diagrams, which emerge from iterations of the
Lippmann-Schwinger equation.} Another complication is related to the
regularization of the Schr\"odinger or Lippmann-Schwinger (LS) equation
\cite{Lepage:1997cs}. Iterations of nuclear potentials in the LS equation generate ultraviolet
divergent higher-order contributions to the amplitude, which cannot be
made finite by counterterms in the truncated potential.  One
is, therefore, forced to keep the
ultraviolet cutoff $\Lambda$ finite (of the order of the breakdown scale
$\Lambda_b$) \cite{Lepage:1997cs,Epelbaum:2009sd,Epelbaum:2018zli}. As discussed in section \ref{sec5}, maintaining
consistency between nuclear forces and currents in the presence of a
finite cutoff is a rather nontrivial task starting from the fourth
order in the chiral expansion.  

In this contribution I will briefly review the current status of
chiral EFT, discuss selected application and address the challenges
that need to be tackled to develop this approach into a precision tool
beyond the two-nucleon system.

\section{Derivation of nuclear forces and currents}
\label{sec2}

As already pointed out in the introduction, nuclear forces and
currents are identified with the irreducible parts of the scattering
amplitude and can be worked out using a variety of methods including
matching to the S-matrix, time-ordered perturbation theory (TOPT) and
the method of unitary transformation (UT). The last approach has been
pioneered in the fifties of the last century in the context of pion
field theory \cite{Fukuda,Okubo:1954zz} and applied to the effective chiral Lagrangian in
Refs.~\cite{Epelbaum:1998ka,Epelbaum:1999dj}. The derivation of
nuclear forces is achieved by 
performing a UT of the effective pion-nucleon Hamiltonian in
the Fock space, which decouples the purely nucleonic
subspace from the rest of the Fock space. The corresponding unitary
operator is determined perturbatively by using the chiral expansion.
The method can be formulated utilizing a diagrammatic language, but
the resulting time-ordered-like graphs have a different meaning
than the ones arising in the context of TOPT. The importance of any
diagram can be estimated by counting
the corresponding power $\nu$ of the chiral expansion parameter $Q \in \{p/\Lambda_b, \;
M_\pi/\Lambda_b\}$, where $M_\pi$ denotes the pion mass and $p \sim
M_\pi$ are three-momenta of the nucleons. For connected diagrams
contributing to the $A$-nucleon potential with $B$ insertions of
external classical sources, the chiral dimension $\nu$ is given 
by \cite{Weinberg:1990rz,Epelbaum:2006eu}
\beq
\label{eq1}
\nu = - 4 - B + 2 (A + L) + \sum _i V_i \Delta_i\,, 
\eeq
where $L$ is the number of loops and $V_i$ is the number of vertices
of dimension $\Delta_i$ which appear in the diagram. The 
dimension of a vertex with $n_i$ nucleon field operators and
$d_i$ derivatives/$M_\pi$-insertions/insertions of the external classical sources is
defined according to \cite{Weinberg:1990rz}
\beq
\label{eq2}
\Delta_i = d_i + \frac{1}{2} n_i - 2 \,.
\eeq
The spontaneously broken chiral symmetry permits only interactions
with $\Delta \geq 0$, so that a finite number of diagrams can be drawn
at each finite chiral order $Q^\nu$. Notice that for actual
calculations in the method of UT, it is more convenient to rewrite
equations (\ref{eq1}), (\ref{eq2}) using different variables as explained
in Ref.~\cite{Epelbaum:2007us}.   

\begin{figure}[tb]
\begin{center}
\includegraphics[width=\textwidth,keepaspectratio,angle=0,clip]{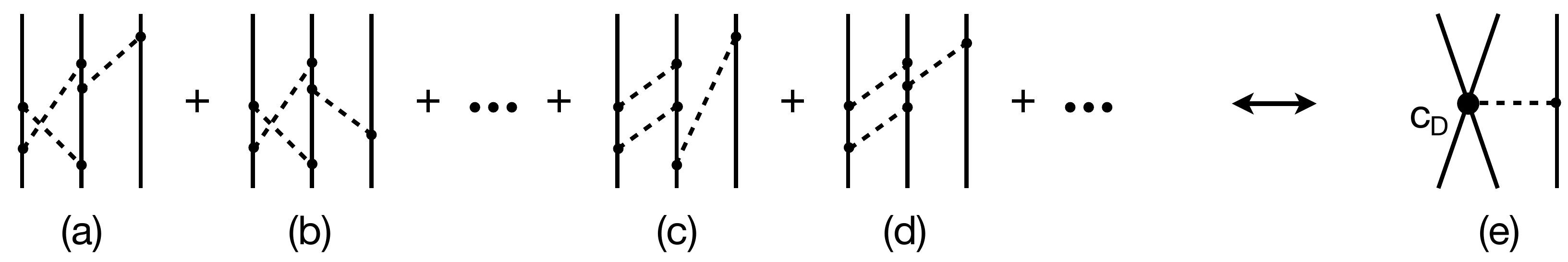}
\end{center}
\vskip -0.5 true cm
    \caption{Time-ordered-like diagrams contributing to the
      two-pion-one-pion-exchange 3NF topology at fourth order in the
      chiral expansion (graphs (a)-(d)) along with the
      one-pion-exchange-contact contribution to the 3NF at third
      order.  Solid and dashed lines refer to nucleons and pions,
      respectively. Solid dots and the filled circle denote vertices
      of dimension $\Delta = 0$ and $\Delta = 1$, respectively.   
\label{fig0}
}
\end{figure}

As already pointed out above, loop contributions to nuclear potentials
can, in general, not be renormalized. The problem is exemplified in Fig.~\ref{fig0} for the
fourth-order (i.e.~N$^3$LO) contribution to the three-nucleon force (3NF)
proportional to $g_A^6$, with $g_A$ referring to the nucleon axial-vector
constant.  
To obtain a renormalized expression for the 3NF, the loop
integrals should only involve linearly divergent pieces, which can be
cancelled by the counterterm in the LEC $c_D$. This is only possible
if the pion exchange between the pair of the first two nucleons and
the last nucleon in diagrams (a)-(d) factorizes out in order to match
the expression for diagram (e) at third chiral order
(i.e.~N$^2$LO). However, evaluating the corresponding diagrams in TOPT,
one finds that the pion exchange does, actually, not factorize out. To ensure
factorization of the one-pion exchange and enable renormalizability of the
3NF\footnote{The situation becomes more complicated if cutoff
  regularization is used instead of dimensional regularization, see
  the discussion in section \ref{sec5}.}, a broad class of additional unitary transformations in the Fock
space has been considered in Refs.~\cite{Epelbaum:2006eu,Epelbaum:2007us}. Other types of
contributions to the 3NF at N$^3$LO and at fifth order (N$^4$LO) in
the chiral expansion and to the current operators show similar
problems with renormalizability. So far, it was always possible to
maintain renormalizability of nuclear forces and current operators, calculated using
dimensional regularization (DR), via a suitable choice of additional
unitary transformations, see Refs.~\cite{Epelbaum:2007us,Kolling:2011mt,Krebs:2012yv,Krebs:2016rqz} for more details. 

\begin{figure}[tb]
\begin{center}
\includegraphics[width=\textwidth,keepaspectratio,angle=0,clip]{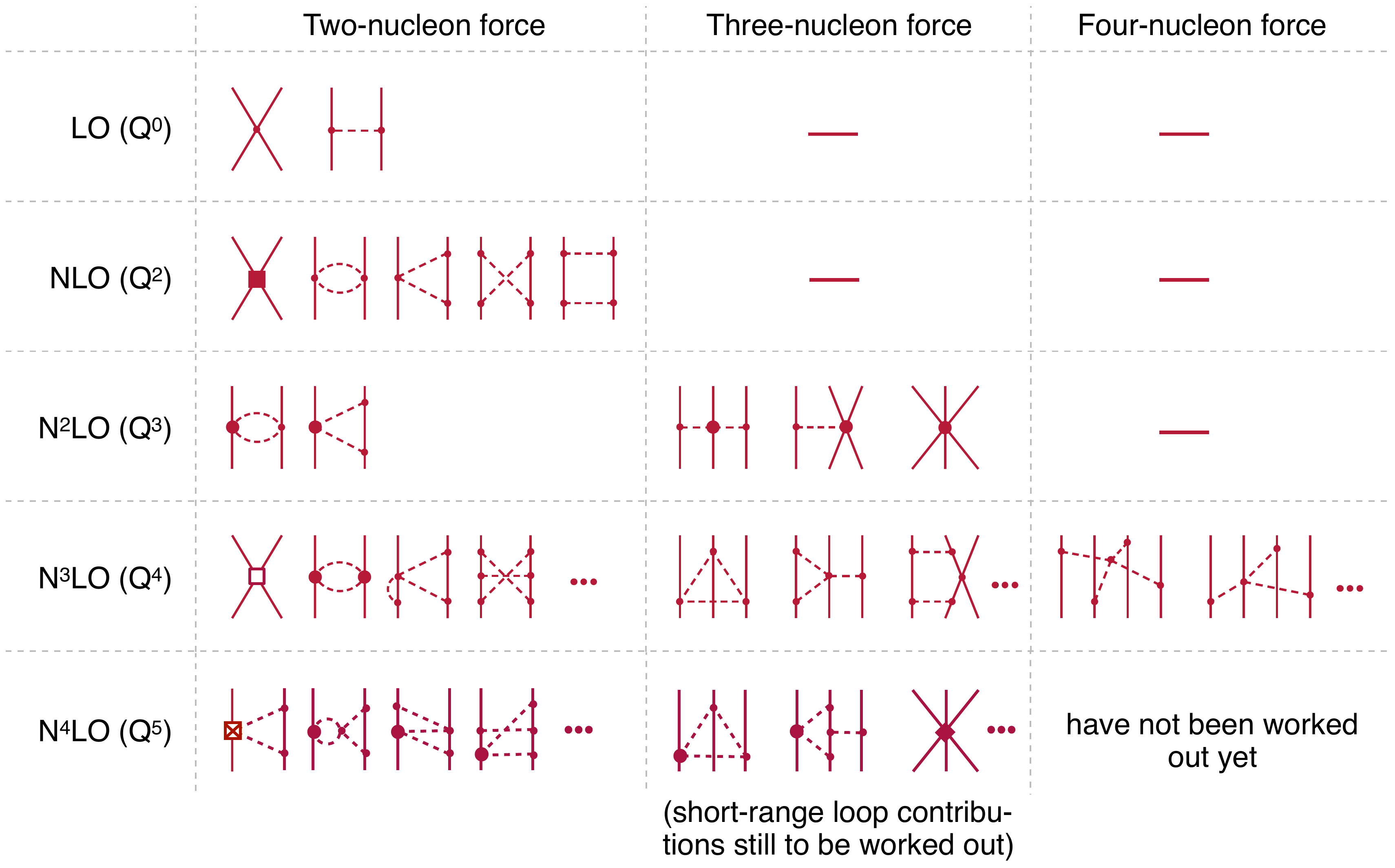}
\end{center}
\vskip -0.5 true cm
    \caption{Chiral expansion of the nuclear forces. Solid and dashed
      lines refer to nucleons and pions. Solid dots, filled circles,
      filled squares, crossed squares and open squares denotes
      vertices from the effective chiral Lagrangian of dimension $\Delta = 0$, $1$, $2$, $3$ and $4$,
      respectively.
\label{fig1}
}
\end{figure}

Fig.~\ref{fig1} visualizes the current state-of-the-art in the
derivation of the nuclear Hamiltonian using the heavy-baryon
formulation of chiral perturbation theory with pions and nucleons as
the only explicit degrees of freedom and utilizing the rules of naive dimensional analysis for
few-nucleon contact operators, see \cite{Nogga:2005hy,Birse:2005um,Valderrama:2016koj} for alternative proposals. 
We remind the reader that all diagrams shown in this and following
figures correspond to irreducible
parts of the scattering amplitude and to be understood as series of
all
possible time-ordered-like graphs for a given 
topology. As already explained before, the precise meaning of these 
diagrams and the resulting contributions to the nuclear forces are
scheme dependent.  

The nucleon-nucleon potential has been calculated to fifth order
(N$^4$LO) in the chiral expansion using dimensional regularization
\cite{Ordonez:1992xp,Friar:1994zz,Kaiser:1997mw,Epelbaum:1998ka,Kaiser:1999ff,Kaiser:1999jg,Kaiser:2001pc,Kaiser:2001at,Entem:2014msa}. 
The expressions for the leading and subleading 3NF can be found in
Refs.~\cite{vanKolck:1994yi,Epelbaum:2002vt,Ishikawa:2007zz,Bernard:2007sp,Bernard:2011zr} and \cite{Epelbaum:2006eu,Epelbaum:2007us}, respectively. Apart from  the contributions
involving NN contact interactions, which still have to be worked out,
the N$^4$LO terms in the 3NF can be found in Refs.~\cite{Krebs:2012yv,Krebs:2013kha,Girlanda:2011fh}. The
leading contribution to the four-nucleon force (4NF) appears at N$^3$LO and
has been derived in Refs.~\cite{Epelbaum:2006eu,Epelbaum:2007us}.
It is important to emphasize that the long-range parts of the
nuclear forces are completely determined by the spontaneously broken
approximate chiral symmetry of QCD along with the experimental
and/or empirical information on the pion-nucleon system needed to
determined the relevant LECs in the effective Lagrangian. In this
sense, the long-range contributions to the nuclear forces and currents
can be regarded as parameter-free predictions. Given that the chiral
expansion of the NN contact operators in the isospin limit contains
only contributions at orders $Q^{2n}$, $n=0,1,2,\ldots$, the N$^2$LO
and the isospin-invariant N$^4$LO corrections to the NN potential are
parameter-free.  This also holds true for the N$^3$LO contributions to
the 3NF and 4NF. For calculations utilizing a formulation of chiral
EFT with explicit $\Delta$(1232) degrees of freedom see
Refs.~\cite{Ordonez:1995rz,Kaiser:1998wa,Krebs:2007rh,Epelbaum:2007sq,Krebs:2018jkc,Piarulli:2014bda,Ekstrom:2017koy}.  

The chiral power counting offers a natural qualitative explanation of
the observed hierarchy of the nuclear forces, and the actual size of the
various contributions to observables generally agrees well with the
expectations based on naive dimensional analysis (NDA) underlying the chiral
power counting. For example, the kinetic energy of the deuteron can
naively be expected of the order of $E_{\rm kin} \sim M_\pi^2/m_N \sim 20$~MeV, since the pion mass
is the only explicit soft scale in the problem. This compares
well with the actual findings of $E_{\rm kin} \simeq 12
\ldots 23$~MeV for the LO$\ldots$N$^4$LO chiral potentials of
Ref.~\cite{Epelbaum:2014efa,Epelbaum:2014sza} and $E_{\rm kin} \simeq 12
\ldots 16$~MeV for the LO$\ldots$N$^4$LO$^+$\footnote{The ``$+$'' sign indicates that
  we have included four contact interactions in F-waves from N$^5$LO
  in order to reproduce several sets of very precisely measured
  proton-proton data at higher energies. The same contact interactions
are also included in the N$^4$LO potentials of Ref.~\cite{Entem:2017gor}.} potentials of
Ref.~\cite{Reinert:2017usi}. For comparison, for the Argonne AV18
potential of \cite{Wiringa:1994wb}, one
finds $E_{\rm kin} \simeq 19$~MeV. For $^3$H, one can use the
following simple considerations to estimate the size of the
3NF. Using phenomenological potentials, one typically finds $\big|
\langle V_{NN} \rangle \big|_{^3H} \sim 50$~MeV.  Thus, according to
the power counting, the 3NF contribution to the $^3$H binding energy
may be expected of the order of  $Q^3 \big|
\langle V_{NN} \rangle \big|_{^3H}$. With $Q  \sim M_\pi/\Lambda_b $
and $\Lambda_b \simeq 600$~MeV \cite{Epelbaum:2014efa}, one expects 
the 3NF to contribute about $0.013 \big|
\langle V_{NN} \rangle \big|_{^3H} \sim  650$~keV to the triton binding energy.  This matches well
with the observed typical underbinding of $^3$H in calculations based
on NN forces only \cite{Binder:2015mbz,Binder:2018pgl}. Similar estimations may be carried out for the 4NF, 
for other light nuclei and for nucleon-deuteron scattering observables. In the latter
case, assuming $Q = \max \{p/\Lambda_b, \;
M_\pi/\Lambda_b\}$, one expects the 3NF effects to be small at low energy,
but become more important at higher energies. Again, this expectation
is in line with the observed discrepancies between nucleon-deuteron
scattering data and calculations based on NN interactions only \cite{Binder:2015mbz,Binder:2018pgl}. 
Notice, however, that the observed
fine tuned nature of
the nuclear force, resulting e.g.~in a small value of the deuteron
binding energy of $E_{\rm b, \, ^2H}
\simeq 2.224$~MeV, cannot be explained by NDA which actually suggests 
$E_{\rm b, \, ^2H} \sim E_{\rm  kin, \, ^2H} \sim \big| \langle V_{NN} \rangle
\big|_{^2H}$. On the other hand, one observes $ \langle V_{NN} \rangle
\big|_{^2H} \simeq - E_{\rm
  kin, \, ^2H}$, so that $E_{\rm b, \, ^2H}  \ll \big| \langle V_{NN} \rangle \big|_{^2H}$. Similar
fine tuning also persists for light nuclei.  

\begin{figure}[tb]
\begin{center}
\includegraphics[width=\textwidth,keepaspectratio,angle=0,clip]{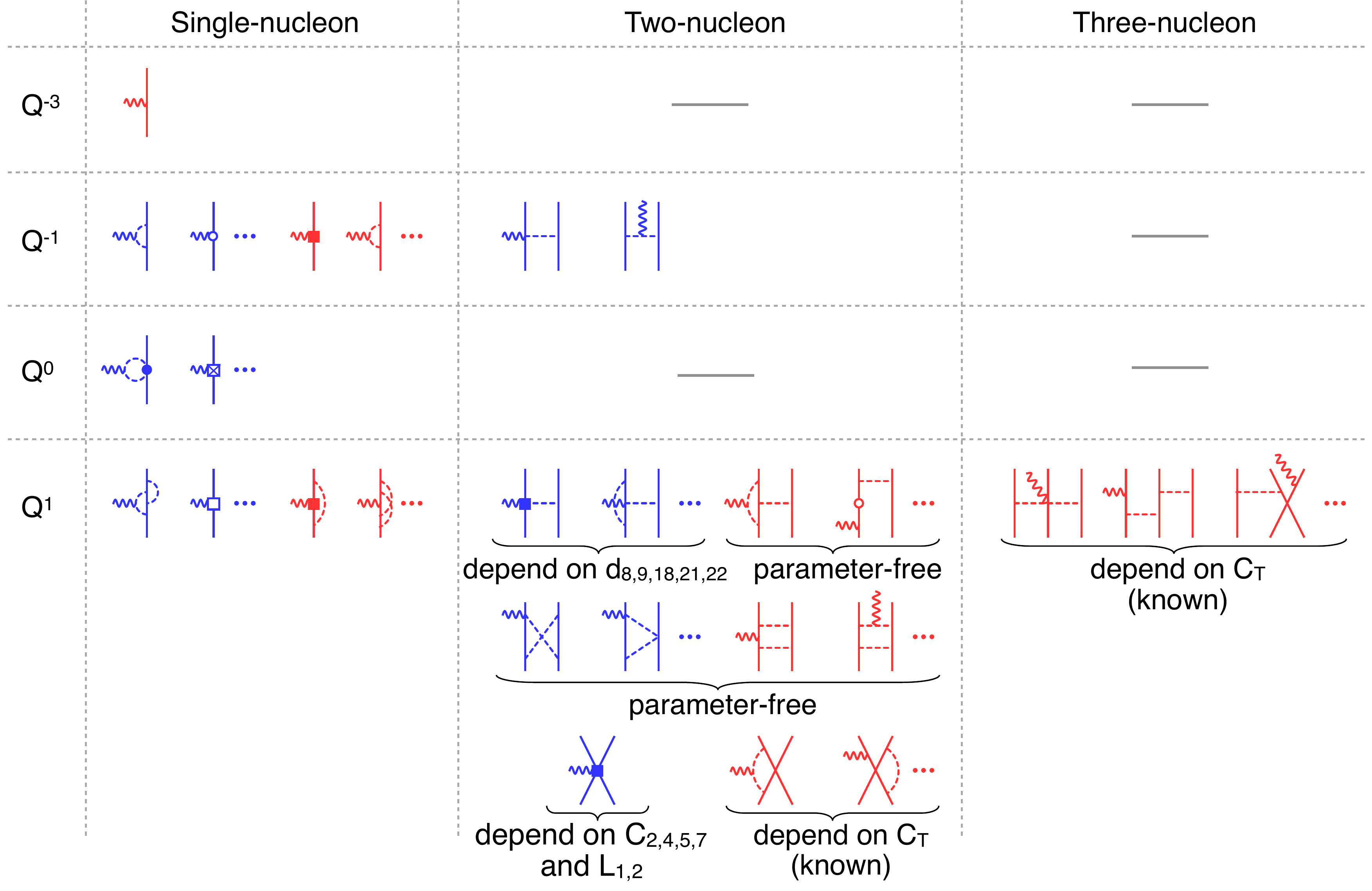}
\end{center}
\vskip -0.5 true cm
    \caption{Chiral expansion of the nuclear electromagnetic
      currents. Red (blue) diagrams show the contributions to the
      charge (current) operators. Wavy lines refer to photons. For
      remaining notations see Fig.~\ref{fig1}. 
\label{fig2}
}
\end{figure}

Nuclear electromagnetic and axial currents have been worked out
in chiral EFT completely up through N$^3$LO.
Fig.~\ref{fig2} summarizes the contributions to the
electromagnetic charge and current operators derived using the method of
UT in Refs.~\cite{Kolling:2009iq,Kolling:2011mt,Krebs:2019aka} and employing DR for loop integrals. 
The resulting expressions are, per construction, off-shell consistent
with the nuclear forces derived by our 
group using the same approach. Again, the hierarchy of the $A$-nucleon contributions to the
charge and current operators suggested by the chiral power counting is
fully in line with empirical findings based on explicit calculations,
which show the dominance of single-nucleon contributions for most
of the low-energy observables \cite{Arenhovel:1990yg,Golak:2005iy}. In particular, the charge operator is
strongly dominated by the one-nucleon (1N) term with the ``meson-exchange'' 
contributions being suppressed by four powers of the expansion
parameter. On the other hand, the power counting suggests that the
three-nucleon (3N) charge operator is as important as the two-body
one, which can be tested e.g.~by calculating elastic form factors (FFs) of light
nuclei. Notice further that up to N$^3$LO, the charge operator does
not involve any unknown LEC. 
It is furthermore important to emphasize that the single-nucleon
contributions to both the charge and current operators can
be expressed in terms of the electromagnetic FFs of
the nucleon. Using the available empirical information on the
nucleon FFs then allows one to avoid relying on their strict chiral
expansion  known to converge
slowly due to large contributions of vector mesons
\cite{Kubis:2000zd,Schindler:2005ke}.

The
leading contributions to the two-nucleon (2N) current operator emerge from a single
pion exchange. The corrections at N$^3$LO include one-loop
contributions to the one- and two-pion exchange as well as short-range
operators. The third-order pion-nucleon LECs $d_{18}$ and $d_{22}$ can be determined  
from the Goldberger-Treiman discrepancy and the axial radius of the
nucleon, while $d_{9, 21, 22}$ contribute to the charged pion
photoproduction and the radiative capture reactions
\cite{Fearing:2000uy,Gasparyan:2010xz}.
For the explicit form of the pion-nucleon Lagrangian see
Ref.~\cite{Fettes:2000gb}. 
The
short-range 2N operators depend, apart from the LECs $C_i$, which
govern the short-range NN potential at N$^2$LO, also on the two new LECs $L_{1,2}$,
which can be determined e.g.~from the deuteron
magnetic moment and the cross section in the process $np \to d \gamma$     
\cite{Chen:1999tn}.

Notice that the expressions for the N$^3$LO contributions to the
electromagnetic charge  and current operators derived in the method of
UT in Refs.~\cite{Kolling:2009iq,Kolling:2011mt,Krebs:2019aka} differ from the ones calculated by Pastore et
al.~using time-ordered perturbation theory \cite{Pastore:2008ui,Pastore:2009is,Pastore:2011ip}. The reader is
referred to Ref.~\cite{Hermann_review} for a comprehensive review article,
which also addresses the differences between the two approaches. 

\begin{figure}[tb]
\begin{center}
\includegraphics[width=\textwidth,keepaspectratio,angle=0,clip]{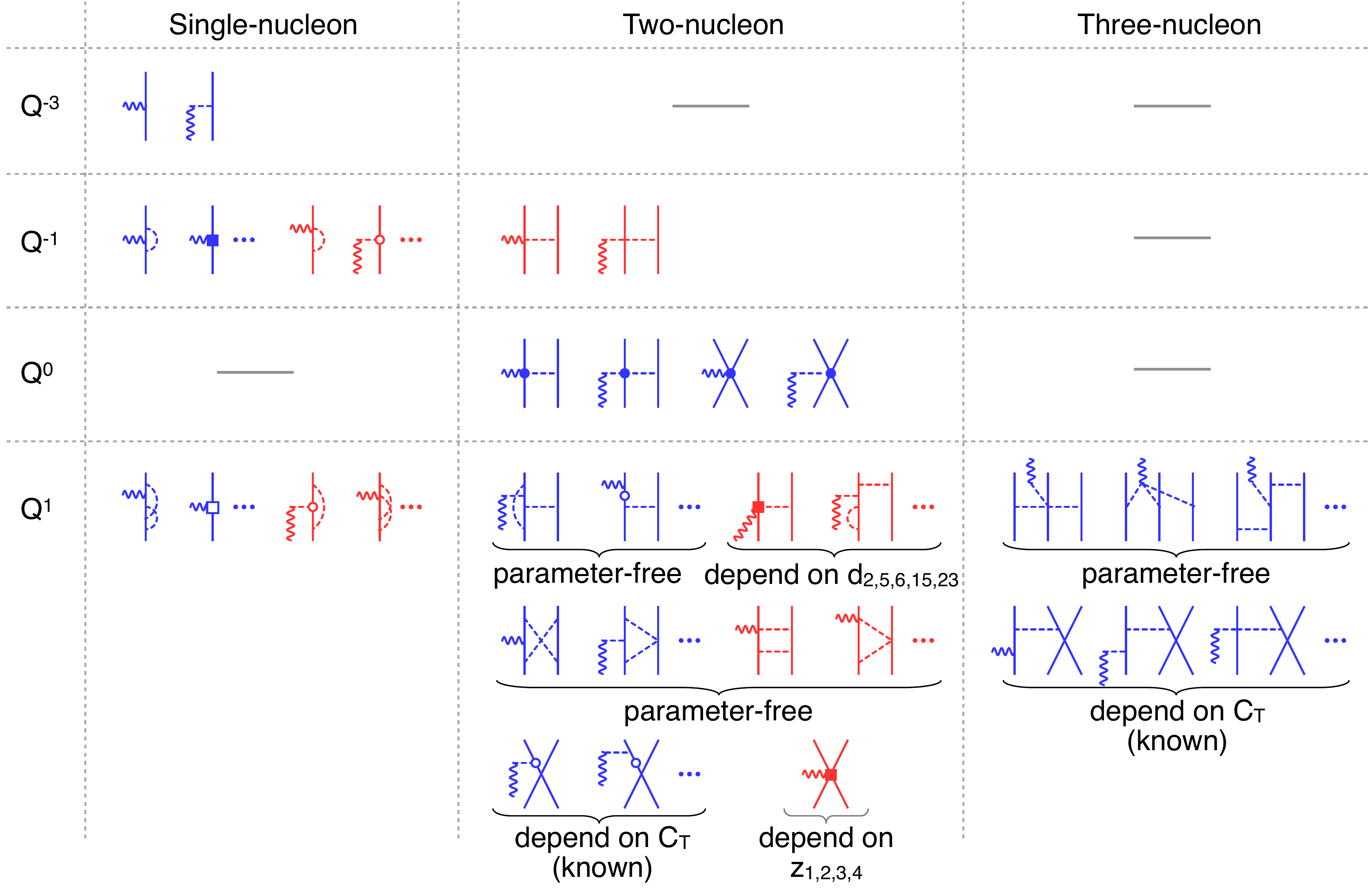}
\end{center}
\vskip -0.5 true cm
    \caption{Chiral expansion of the nuclear axial 
      currents. Red (blue) diagrams show the contributions to the
      charge (current) operators. Wavy lines refer to external
      axial-vector sources. For
      remaining notations see Fig.~\ref{fig1}. 
\label{fig3}
}
\end{figure}

Recently, these studies have been extended to derive the nuclear axial
and pseudoscalar currents up to N$^3$LO  using
the method of UT \cite{Krebs:2016rqz}, see Fig.~\ref{fig3}. 
Interestingly, one observes exactly the
opposite pattern as compared to the electromagnetic operators with
the dominant contributions to the 1N current, 2N charge and 3N current
operators appearing at LO, NLO and N$^3$LO, respectively. In a
complete analogy with the electromagnetic currents, the 1N
contributions are expressible in terms of the corresponding FFs of the
nucleon. The 2N and 3N contributions to the current density are
parameter free at this order, while the long-range one-loop corrections to the 2N charge density
depend on a number of poorly known LECs $d_i$. In addition, there are
four new short-range operators contributing to the 2N charge operator
at N$^3$LO. Again, we emphasize that our results deviate from the (incomplete) calculation
by Baroni et al.~\cite{Baroni:2015uza} using time-ordered perturbation
theory and refer the reader to Ref.~\cite{Hermann_review} for a detailed
comparison. 

\section{High-precision chiral two-nucleon potentials}
\label{sec3}

In the last section I briefly reviewed the state of the art in the
derivation of the nuclear forces and currents. These calculations are
carried out using DR to regularize divergent loop integrals. As
pointed out in the introduction, the derived nuclear potentials and
current operators are singular at short distances and need to be
regularized. To the best of my knowledge, it is not known how to
subtract all ultraviolet divergences arising from iterations of the NN
potential in the Lippmann-Schwinger equation. Thus, the cutoff
$\Lambda$ has to be kept finite of the order of the breakdown
scale $\Lambda_b$
\cite{Lepage:1997cs,Epelbaum:2009sd,Epelbaum:2018zli}.
In Ref.~\cite{Epelbaum:2014efa}, this scale was estimated to be
$\Lambda_b \sim 600$~MeV. This was confirmed in the Bayesian analysis
of Ref.~\cite{Furnstahl:2015rha}, which found that it may even be 
somewhat larger, see \cite{Melendez:2017phj,Wesolowski:2018lzj} for a related
recent work along this line. In practice, even lower values of the
cutoff $\Lambda$ are preferred in order to avoid the appearance of
deeply bound states, which would complicate the numerical
treatment of the nuclear $A$-body problem, and to keep the potentials
sufficiently soft in order to facilitate the convergence of many-body
methods. It is, therefore, important to use a regulator, which
minimizes the amount of finite-cutoff artifacts. In Ref.~\cite{Epelbaum:2014efa}, we
argued that a local regularization of the pion-exchange
contributions to the nuclear forces is advantageous compared to
the nonlocal regulators used e.g.~in \cite{Epelbaum:1999dj,Epelbaum:2004fk,Entem:2003ft} as it maintains the
analytic properties of the potential and does not induce long-range
artifacts, see also Refs.~\cite{Gezerlis:2014zia,Piarulli:2014bda} for a related discussion. 
In Refs.~\cite{Epelbaum:2014efa,Epelbaum:2014sza}, a coordinate-space cutoff was employed to
regularize the one- and two-pion exchange contributions. However,
the implementation of such a regulator in coordinate space has turned
out to be 
technically difficult for the 3NF and current operators. Thus, in
Ref.~\cite{Reinert:2017usi}, a momentum-space version of the local regulator was
introduced by replacing the static propagators of pions exchanged
between different nucleons via $(\vec q \, ^2 + M_\pi^2)^{-1} \;
\longrightarrow \; (\vec q \, ^2 + M_\pi^2) ^{-1}\; \exp\big[ -(\vec q \, ^2
+ M_\pi^2)/\Lambda^2 \big]$, see Ref.~\cite{Rijken:1990qs} for a similar approach.  
Obviously, the employed regulator does not induce long-range artifacts
at any finite order in the $1/\Lambda$-expansion. Notice further that
the long-range part of the two-pion exchange NN potential, derived
using DR, does not need to be recalculated using the new regulator. As
shown in  \cite{Reinert:2017usi}, regularization of the two-pion
exchange can be easily
accounted for using the spectral-function representation.
For example, for a central two-pion exchange potential $V (q)$,
regularization is achieved via
\beq
V (q) = \frac{2}{\pi} \int_{2 M_\pi}^\infty \mu d \mu \frac{\rho (\mu
  )}{\vec q \, ^2 + \mu^2} + \ldots \; \longrightarrow \;    \frac{2}{\pi} \int_{2 M_\pi}^\infty \mu d \mu \frac{\rho (\mu
  )}{\vec q \, ^2 + \mu^2} e^{- \frac{\vec q \, ^2 + \mu^2}{2 \Lambda^2}}+ \ldots \,,
\eeq
where $\rho (\mu )$ is the corresponding spectral function and the
ellipses refer to the contributions polynomial in $\vec q \, ^2$ and
$M_\pi$. In addition, a final number of (locally regularized)
subtraction terms allowed by the power counting are taken into account
to ensure that the corresponding long-range potentials and
derivatives thereof vanish at the origin. For the contact NN
interactions, a simple non-local cutoff of the Gaussian type
$\exp\big[ -(\vec p \, ^2 
+ \vec p^{\, \prime} {}^2)/\Lambda^2 \big]$ with $\vec p$ and
$\vec p^{\, \prime}$ denoting the initial and final center-of-mass
momenta was employed. Using  
this regularization
scheme and adopting the pion-nucleon LECs from the recent analysis in
the framework of the Roy-Steiner equations \cite{Hoferichter:2015tha,Hoferichter:2015hva}, a family of
new chiral NN potentials from LO to N$^4$LO$^+$ was presented in
Ref.~\cite{Reinert:2017usi} for the cutoff values of $\Lambda =
\big\{350, \; 400, \; 450, \; 500, \; 550\}$~MeV.  
The resulting potentials at N$^4$LO$^+$ are currently the most precise chiral
EFT NN potentials on the market. For the medium cutoff choice of
$\Lambda = 450$~MeV, the description of the neutron-proton and
proton-proton scattering data from the 2013 Granada database \cite{Perez:2013jpa}
below $E_{\rm lab} = 300$~MeV is essentially perfect at N$^4$LO$^+$ as witnessed by the
corresponding $\chi^2$ values of $\chi^2/\mbox{datum} = 1.06$ and    
$\chi^2/\mbox{datum} = 1.00$, respectively. The N$^4$LO$^+$ potentials
of Ref.~\cite{Reinert:2017usi} thus qualify to be regarded as partial
wave analysis (PWA). Distinct features of these potentials in comparison
with the other available chiral EFT interactions are summarized in
Ref.~\cite{CD18}.

\begin{figure}[tb]
\begin{center}
\includegraphics[width=0.975\textwidth,keepaspectratio,angle=0,clip]{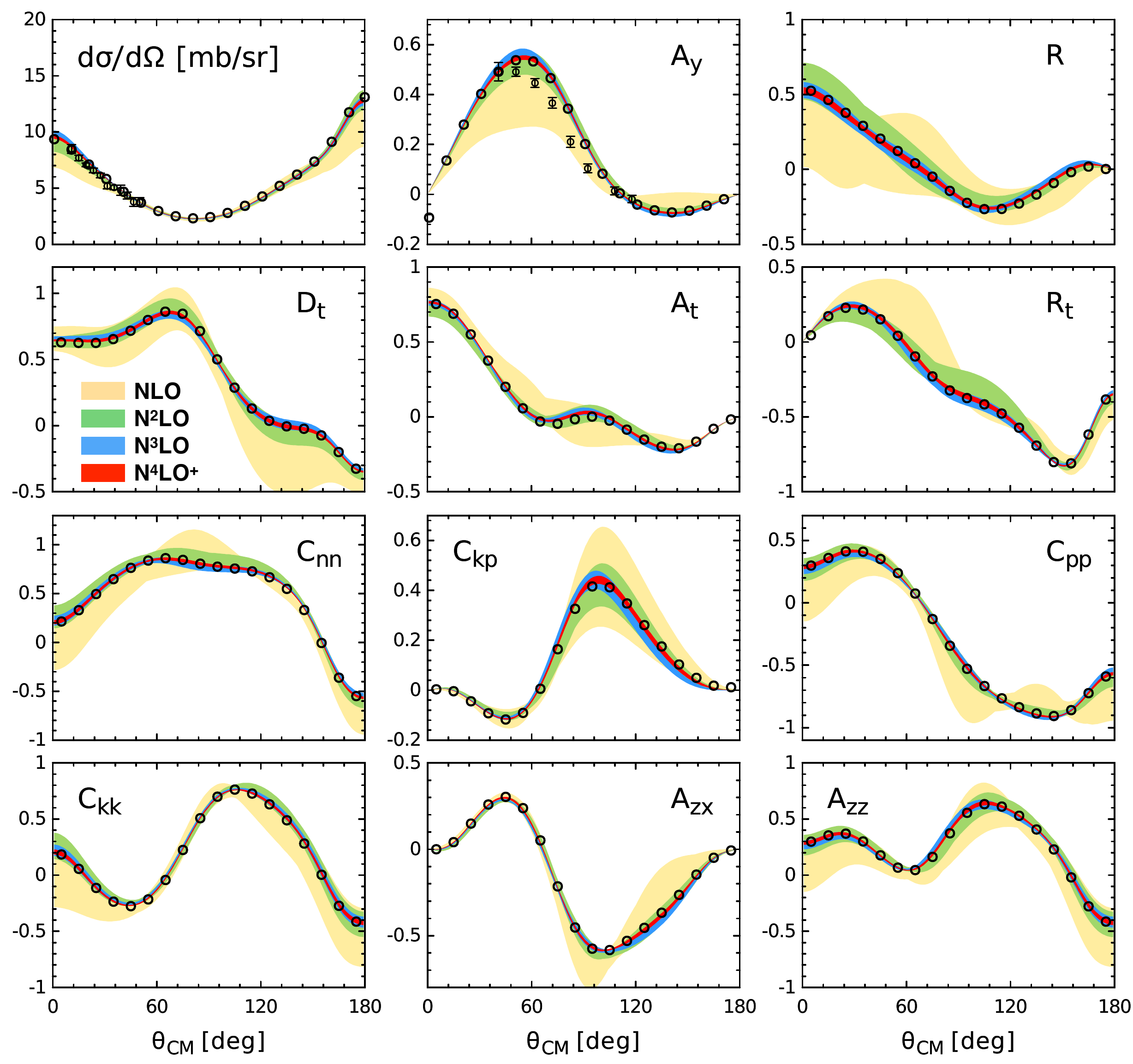}
\end{center}
\vskip -0.5 true cm
\caption{Neutron-proton scattering observables 
at  $E_{\rm   lab} = 143\,$MeV calculated up to N$^4$LO$^+$ using the
chiral NN potentials of Ref.~\cite{Reinert:2017usi} for the 
      cutoff of $\Lambda=450\,$MeV.  Data for the cross section are at
      $E_{\rm lab} = 142.8\,$MeV and taken from \cite{ber76} and for
      the analyzing power $A_y$ 
from \cite{kuc61}. Bands show the estimated truncation error while  open circles are the results of the Nijmegen
PWA \cite{Stoks:1993tb}. 
\label{fig4}
}
\end{figure}

As a representative example, we show in Fig.~\ref{fig4} the
description of the neutron-proton scattering observables at  $E_{\rm lab} \simeq 143$~MeV
at various orders of the chiral expansion.
The truncation bands have been generated using the algorithm
formulated in Ref.~\cite{Epelbaum:2014efa}. For the application of the
Bayesian approach for the quantification of truncation errors to the
potentials of Ref.~\cite{Reinert:2017usi} see Ref.~\cite{CD18}.
One observes excellent 
convergence of the chiral expansion and a very good agreement with 
the Nijmegen PWA. These conclusions also hold true for other
scattering observables and deuteron properties.  

\section{The three-nucleon force}
\label{sec4}

The novel semi-local chiral NN potentials of Refs.~\cite{Epelbaum:2014efa,Epelbaum:2014sza,Reinert:2017usi} have
already been explored in nucleon-deuteron scattering and selected
nuclei
\cite{Binder:2015mbz,Binder:2018pgl,Skibinski:2016dve,Witala:2016inj,Witala:2019ffj}. By
calculating various few-nucleon observables using 
the NN interactions only, a clear discrepancies between experimental
data and theoretical results well outside the
range of the estimated uncertainties were observed. The magnitude of
these discrepancies appears to be consistent with the expected size
of the 3NF, which start contributing at N$^2$LO, see Fig.~\ref{fig1}. 

\begin{figure}[tb]
\begin{center}
\includegraphics[width=\textwidth,keepaspectratio,angle=0,clip]{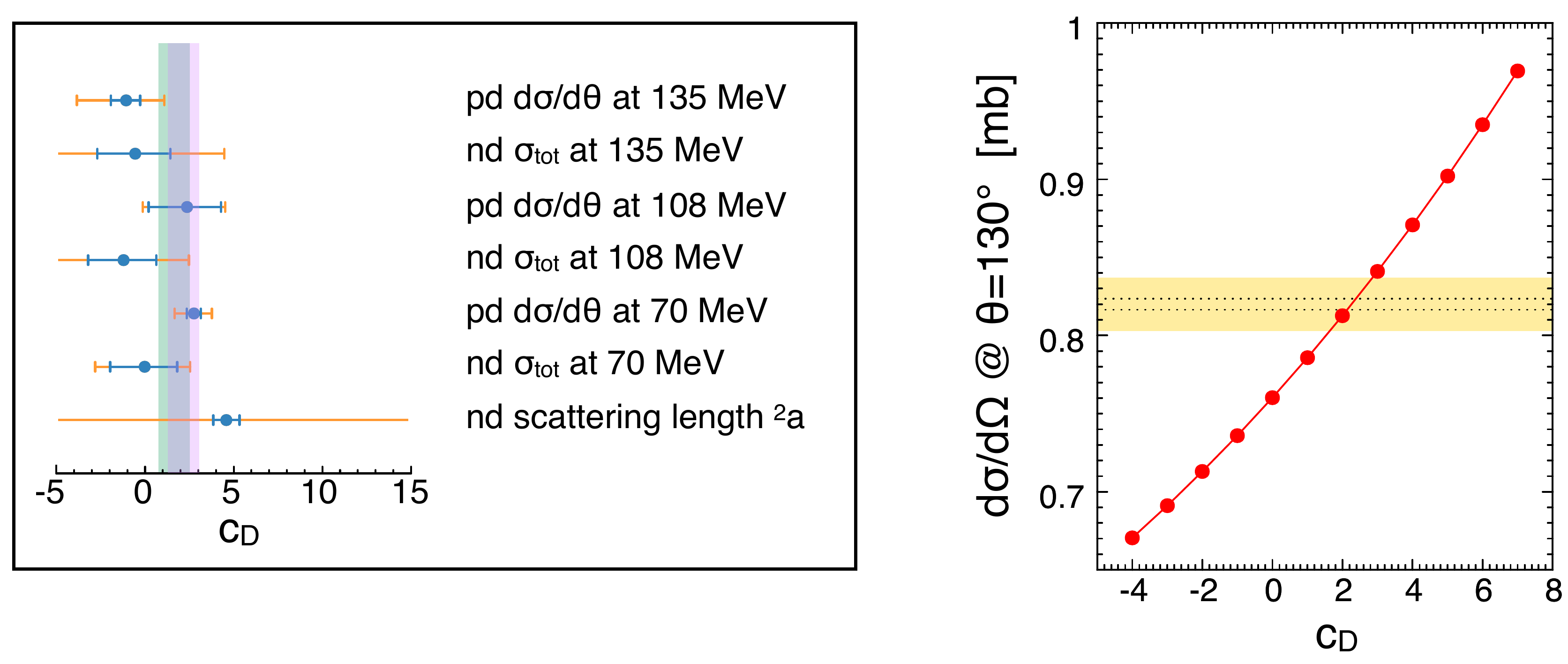}
\end{center}
\vskip -0.5 true cm
\caption{Left: Determination of the LEC $c_D$ from various Nd
  scattering observables as explained in the text for the
  coordinate-space cutoff $R=0.9$~fm.   The smaller (blue) error bars
  correspond to the experimental uncertainty while the larger (orange)
  error bars also take into account the truncation error at
  N$^2$LO. Right: Nd cross section in the minimum region
  ($\theta = 130^\circ$) at $E_N = 70$~MeV as function of the LEC $c_D$. For
  each $c_D$ value, the LEC $c_E$ is adjusted to the $^3$H binding
  energy. Dotted lines show the statistical uncertainty of the
  experimental data from Ref.~\cite{Sekiguchi:2002sf}, while the yellow band also
  takes into account the quoted systematic uncertainty of $2\%$.  
\label{fig5}
}
\end{figure}

To perform complete calculations at N$^2$LO and beyond one needs to
include the 3NF (and 4NF starting from N$^3$LO). These have to be
regularized in a way {\it consistent} with the NN potentials.  The
precise meaning of {\it consistency} in this context will be
addressed in the next section. In Ref.~\cite{Epelbaum:2018ogq}, we performed
calculations of nucleon-deuteron (Nd) scattering and the ground and
low-lying excited states of light nuclei up to $A = 16$  up through
N$^2$LO using the semilocal coordinate-space regularized NN potentials
of Refs.~\cite{Epelbaum:2014efa}  together with the 3NF regularized in the same
way. Notice that the Faddeev equations are usually solved
in the partial wave basis. Partial wave decomposition of
arbitrary 3NFs can be accomplished numerically using the machinery
developed in Refs.~\cite{Golak:2009ri,Hebeler:2015wxa}. The leading
3NF at N$^2$LO depends on two LECs $c_D$ and $c_E$, which cannot be
determined in the NN system. It is customary to tune the short-range
part of the 3NF in such a way that the $^3$H and/or $^3$He binding
energies (BEs) are reproduced. As for the second constraint, different
options have been explored including the neutron-deuteron spin-$1/2$
scattering length $^2a$, the BE and/or radius of the
$\alpha$-particle, Nd scattering observables, selected
properties of light and medium-mass nuclei, equation of state for
symmetric nuclear matter, see Ref.~\cite{Hammer:2012id} for a review. 
In Ref.~\cite{Epelbaum:2018ogq}, we have explored the possibility to
determine both LECs entirely from the three-nucleon
system. Specifically, we used the triton BE to express
$c_E$ as a function of $c_D$. To fix the $c_D$ value, a range of the
available differential and total cross section data in elastic Nd
scattering and the doublet scattering length were considered. Notice
that the 3NF is well known to have a large impact on the differential
cross section in the minimum region (at not too low energies)
\cite{Gloeckle:1995jg}. Taking into account the estimated theoretical uncertainty
from the truncation of the chiral expansion, the very precise
experimental data of Ref.~\cite{Sekiguchi:2002sf} for the proton-deuteron differential
cross section at $E_{\rm N} = 70$~MeV were found to impose the strongest constraint on the
$c_D$ value as visualized in Fig.~\ref{fig5}
It is important to emphasize that contrary to the scattering length
$^2a$ and the $^4$He BE, we do not observe any
correlations between the triton BE and the cross section minimum in
elastic Nd scattering at the considered energies. In particular, we
found that a variation of the $^3$H BE used in the fit effects the
value of $c_E$ but has almost no effect on the value of $c_D$. 
Having determined the LECs $c_D$ and $c_E$ as described above, the
resulting nuclear Hamiltonian was used to calculate selected Nd
scattering observables and low-lying states in nuclei up to $A =
16$. The inclusion of the 3NF was found to improve the agreement with
the data for most of the considered observables.  

While these results are quite promising, the theoretical uncertainty
of the N$^2$LO approximation is still fairly large
\cite{Epelbaum:2018ogq}. At higher chiral orders, the estimated
truncation errors are, however, expected to become 
significantly smaller than observed deviations between experimental data and
theoretical calculations. This is especially true for Nd scattering
observables at intermediate energies, see Ref.~\cite{Binder:2015mbz,Binder:2018pgl} for
examples. Notice that Nd scattering observables are known to be  not
very sensitive to the off-shell behavior of NN interactions as
shown by Faddeev calculations using a variety of
essentially phase equivalent, high-precision phenomenological potentials \cite{Gloeckle:1995jg}. This feature
persists for high-precision chiral NN potentials at
N$^4$LO$^+$. The large discrepancies between theory and data for spin
observables in Nd scattering \cite{KalantarNayestanaki:2011wz}, therefore, seem to be universal
and should presumably be attributed to the deficiencies of the available 3NF
models. 3NF effects at N$^2$LO appear to be qualitatively similar to
the ones of the phenomenological models such as the
Tucson-Melbourne \cite{Coon:2001pv} or Urbana IX \cite{Pudliner:1997ck} 3NFs and are
insufficient to resolve the above mentioned discrepancies.
The solution of the long-standing 3NF challenge is therefore likely to
emerge from corrections to the 3NF beyond N$^2$LO. Based on the
experience in the NN sector \cite{Reinert:2017usi}, the description of Nd scattering
data will likely require pushing the chiral expansion to (at least)
N$^4$LO. 

\section{Towards consistent regularization of nuclear forces and currents}
\label{sec5}

To take into account the chiral 3NF in few-body calculations, the
resulting expressions, derived using DR as discussed in section \ref{sec2}, have to be regularized in
the way {\it consistent} with the NN force. As will be explained below, this
poses a nontrivial problem starting from N$^3$LO, where the first
loop contributions appear in the 3NF. 

\begin{figure}[tb]
\begin{center}
\includegraphics[width=\textwidth,keepaspectratio,angle=0,clip]{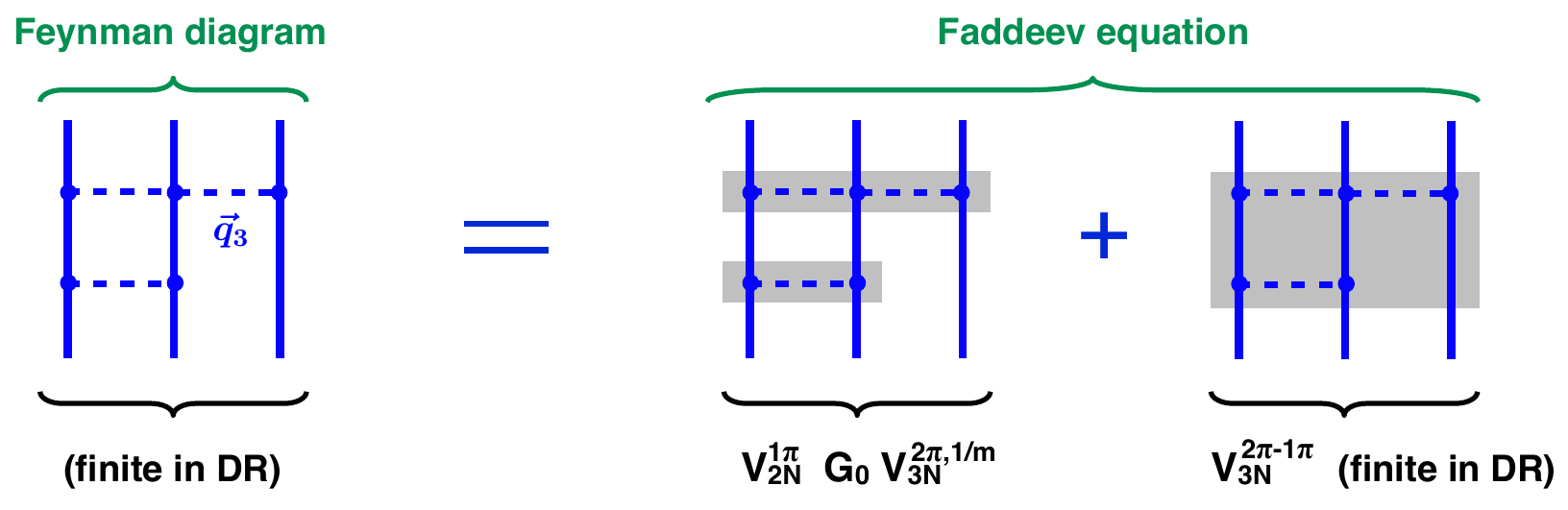}
\end{center}
\vskip -0.5 true cm
\caption{The on-shell amplitude from the one-pion-two-pion-exchange
  Feynman diagram (left) is represented as in terms of iterations of the Faddeev
  equation (right). Gray-shaded rectangles visualize the corresponding
  two- and three-nucleon potentials.   
\label{fig6}
}
\end{figure}

As already pointed out above, nuclear potentials and currents are not uniquely
defined due to inherent unitary ambiguities. Off-shell behaviors of the 2NF
and 3NF must be consistent with each other in order to ensure that
iterations of the Lippmann-Schwinger (or Faddeev) equations reproduce
the corresponding on-shell contributions to the S-matrix (up to 
higher-order corrections), as exemplified in Fig.~\ref{fig6} for one
particular contribution to the 3N scattering amplitude. 
As already mentioned before, all expressions for the nuclear forces and current
operators reviewed in section \ref{sec2}, which are derived using the
method of UT and employing DR to regularize divergent loop integrals, are
consistent with each other {\it provided one also uses DR to regularize loops
from iterations of the integral equation}, see e.g.~the
first diagram on the right-hand side of the equation in
Fig.~\ref{fig6}. However, in practice, regularization of the $A$-body Schr\"odinger
equation in the context of nuclear chiral EFT is achieved by
introducing a cutoff rather than by using DR.
This raises the important question of whether the usage of nuclear potentials,
derived in DR and subsequently regularized with a cutoff, still yields
results which are consistent in the above-mentioned sense. It is easy
to see that this is, generally, not the case by looking at the example
shown in Fig.~\ref{fig6}. The expressions for the
two-pion-exchange 3NF proportional to $g_A^2/m_N$ in the second
diagram are given in Eqs.~(4.9)-(4.11) 
of Ref.~\cite{Bernard:2011zr}, while  the expression for the 
two-pion-one-pion
exchange 3NF proportional to $g_A^4$  in the last diagram, evaluated
in DR, can be found in Eqs.~(2.16)-(2.20) of 
Ref.~\cite{Bernard:2007sp}. The one-pion exchange potential
regularized with a local momentum-space cutoff discussed in section
\ref{sec3} is given in Ref.~\cite{Reinert:2017usi}. Using the same
regulator for the two-pion exchange 3NF, one finds that the
$V_{2N}^{1 \pi} \, G_0 \, V_{3N}^{2 \pi, \, 1/m}$ contribution
in Fig.~\ref{fig6} contains linear divergent terms of the
kind
\beq
\label{temp1}
\sim \Lambda \frac{q_1^i q_3^j}{\vec q_3^{\,2} + M_\pi^2}, \quad \quad
\sim \Lambda \frac{q_3^i q_3^j}{\vec q_3^{\, 2} + M_\pi^2}\,.
\eeq
While the last divergence can be absorbed into the LEC $c_D$, the
first divergent term violates the chiral symmetry, since it corresponds
to a derivative-less coupling of the exchanged pion. No such
chiral-symmetry-breaking contribution appears in the 3NF at N$^2$LO. 
On the other hand, the Feynman diagram on the left-hand-side of the
equation in Fig.~\ref{fig6} must, of course, be renormalizable not only in DR but
also using cutoff regularization (provided it respects the chiral
symmetry). The issue with the renormalization on the right-hand side of
this equation is actually caused by using different regularization schemes when calculating the
reducible (i.e.~iterative) and irreducible contributions to the
amplitude. Re-calculating $V_{3N}^{2\pi-1\pi}$ using the cutoff
regularization instead of DR yields a linearly-divergent contribution,
which cancels exactly the problematic divergence given by the first
term in Eq.~(\ref{temp1}), and restores renormalizability of the scattering amplitude (and
consistency). This example shows that a naive cutoff regularization of the
3NFs, derived using DR, is, in fact, inconsistent starting from
N$^3$LO. Similar problems appear in calculations involving exchange
currents, see Ref.~\cite{CD18-Hermann} for a discussion and an explicit example. 

It is important to emphasize that the above-mentioned problems do not
affect calculations in the NN sector (for the physical value of the
quark masses). This is because the chiral symmetry does not impose constraints
on the momentum dependence of the NN contact interactions. It is,
therefore, always possible to absorb all appearing ultraviolet
divergences into redefinition of the corresponding LECs. 
This is different for contact interactions involving pion fields,
which are indeed
strongly constrained by the spontaneously broken chiral symmetry of
QCD.   

Last but not least, it is important to keep in mind that introducing a cutoff in the way
compatible with the chiral and gauge symmetries is a rather nontrivial
problem. The so-called higher-derivative regularization introduced by
Slavnov in \cite{Slavnov:1971aw} provides one possibility to implement a cutoff in
the chirally invariant fashion already at the level of the effective
Lagrangian, see
Refs.~\cite{Djukanovic:2004px,Djukanovic:2007zz,Behrendt:2016nql} for
applications in the context of chiral EFT. 

\section{Summary}

There have been remarkable progress in pushing chiral EFT into a
precision tool. This theoretically well-founded approach is firmly
rooted in the symmetries of QCD and their breaking pattern.
It allows one to address various low-energy hadronic reactions involving
pions, nucleons and external sources in a systematically improvable
fashion within a unified framework, thus putting nuclear physics onto a
solid basis. 

In this contribution I focused mainly on the applications of chiral EFT
in the few-nucleon sector. During the past two and a half decades,
two-nucleon forces have been worked out
completely up through N$^4$LO while the expressions for the
3NF, 4NF as well as the nuclear electromagnetic and axial currents are
currently available at N$^3$LO. The last generation chiral NN
potentials of Refs.~\cite{Reinert:2017usi,Entem:2017gor} benefit from
the recent analysis of pion-nucleon scattering
in the framework of the Roy-Steiner equation \cite{Hoferichter:2015tha}, which
allows one to reconstruct the long-range part of the nuclear force in
a parameter-free way. The resulting N$^4$LO$^+$ potentials of
Ref.~\cite{Reinert:2017usi} reach the same or even better quality in reproducing NN
scattering data below the pion production threshold
as the phenomenological high-precision potentials, but have 
$\sim 40\%$ less adjustable parameters. This reduction signifies
the importance of the two-pion exchange, which is completely determined
by the spontaneous chiral symmetry of QCD along with the empirical
information on the pion-nucleon system. Another important development
concerns establishing a simple and reliable approach for estimating
truncation errors
\cite{Epelbaum:2014efa,Epelbaum:2014sza,Furnstahl:2015rha,Melendez:2017phj,Wesolowski:2018lzj},
which usually dominate the error budget in 
chiral EFT,  and exploring the other sources of uncertainties \cite{Reinert:2017usi,Ekstrom:2014dxa,Skibinski:2018dot}. 

These developments in the NN sector provide a solid basis for
applications to heavier systems and/or processes involving electroweak
probes. In contrast to the NN force, the 3NFs are still poorly
understood, and the large discrepancies between theory and data in the
three-nucleon continuum pose a long-standing challenge in nuclear
physics \cite{KalantarNayestanaki:2011wz}. While the leading 3NF at N$^2$LO has already been extensively
investigated in Nd scattering and nuclear structure calculations and
demonstrated to yield promising results, it
is certainly insufficient to resolve the observed discrepancies. To
include higher-order contributions to the 3NF (and the nuclear
electroweak currents beyond N$^2$LO), one needs to
introduce a regulator in the way {\it consistent} with the 2NF,
which poses a nontrivial problem starting from N$^3$LO. Work
along these lines is in progress. Another challenge that will have to
be overcome is the
determination of the LECs accompanying short-range contributions of
the 3NF at N$^4$LO, see Ref.~\cite{Girlanda:2018xrw} for an exploratory study.

\vfill 
\section*{Acknowledgments}

It is a great pleasure to thank all my collaborators, and especially
Hermann Krebs and Ulf-G.~Mei{\ss}ner,  for sharing their insights into
various topics discussed in this contribution and useful comments on
the manuscript. I also thank the
organizers of NTSE-2018 for making this interesting and stimulating
workshop possible. 
This work was supported in part by DFG (SFB/TR 110, ``Symmetries and the Emergence of
Structure in QCD'') and the BMBF (Grant No.~05P15PCFN1).

\end{document}